\pdfoutput=1
\documentclass[twocolumn,twoside,slac_two,superscriptaddress]{revtex4}
\usepackage{graphicx}
\usepackage{fancyhdr}
\usepackage{hyperref}
\hypersetup{
    colorlinks=true,
    urlcolor=magenta,
}
\begin{document}

\title{Development of a SiPM Cherenkov camera demonstrator for the CTA observatory telescopes}

%

%

\author{M. Ambrosio}
\affiliation{Universit\`a degli studi di Napoli, Italy}

\author{G. Ambrosi}
\affiliation{INFN Perugia, Italy}

\author{C. Aramo}
\affiliation{Universit\`a degli studi di Napoli, Italy}

\author{E. Bissaldi}
\affiliation{Universit\`a e Politecnico di Bari, Italy}
\affiliation{INFN Bari, Italy}

\author{A. Boiano}
\affiliation{Universit\`a degli studi di Napoli, Italy}

\author{C. Bonavolont\`a}
\affiliation{Universit\`a degli studi di Napoli, Italy}
\affiliation{INFN Napoli, Italy}

\author{E. Fiandrini}
\affiliation{INFN Perugia, Italy}
\affiliation{University of Perugia, Italy}

\author{N. Giglietto}
\affiliation{Universit\`a e Politecnico di Bari, Italy}
\affiliation{INFN Bari, Italy}

\author{F. Giordano}
\affiliation{Universit\`a e Politecnico di Bari, Italy}
\affiliation{INFN Bari, Italy}

\author{M. Ionica}
\affiliation{INFN Perugia, Italy}

\author{C. de Lisio}
\affiliation{Universit\`a degli studi di Napoli, Italy}
\affiliation{INFN Napoli, Italy}

\author{V. Masone}
\affiliation{INFN Napoli, Italy}

\author{R. Paoletti}
\affiliation{Universit\`a di Siena, Italy}
\affiliation{INFN Pisa, Italy}

\author{V. Postolache}
\affiliation{INFN Perugia, Italy}

\author{A. Rugliancich}
\affiliation{Universit\`a di Siena, Italy}
\affiliation{INFN Pisa, Italy}

\author{D. Simone}
\affiliation{INFN Bari, Italy}

\author{V. Vagelli}\email[Corresponding author: ]{valerio.vagelli@pg.infn.it}
\affiliation{INFN Perugia, Italy}
\affiliation{University of Perugia, Italy}

\author{M. Valentino}
\affiliation{CNR-Spin Napoli, Italy}
\affiliation{INFN Napoli, Italy}

\author{L. di Venere}
\affiliation{Universit\`a e Politecnico di Bari, Italy}
\affiliation{INFN Bari, Italy}

\begin{abstract}
The Cherenkov Telescope Array (CTA) Consortium is developing the new generation of ground observatories for the detection of ultra-high energy gamma-rays.
The Italian Institute of Nuclear Physics (INFN) is participating to the R\&D of a possible solution for the Cherenkov photon cameras based on Silicon Photomultiplier (SiPM) detectors sensitive to Near Ultraviolet (NUV) energies.
The latest NUV-HD SiPM technology achieved by the collaboration of INFN with Fondazione Bruno Kessler (FBK) is based on $30\mu\mbox{m}\times30\mu\mbox{m}$ micro-cell sensors arranged in a $6\times6\;\mbox{mm}^2$ area. Single SiPMs produced by FBK have been tested and their performances have been found to be suitable to equip the CTA cameras.
Currently, INFN is developing the concept, mechanics and electronics for prototype modules made of 64 NUV-HD SiPMs intended to equip a possible update of the CTA Prototype Schwarzschild-Couder Telescope (pSCT) telescope. The performances of NUV-HD SiPMs and the design and tests of multi-SiPM modules are reviewed in this contribution.
\end{abstract}

\maketitle

\thispagestyle{fancy}

The analysis of the properties of high energy cosmic ray photons with energies above 100\,GeV provide important information for the understanding of the high energy astrophysical phenomena and for investigating unsolved problems of particle physics, like the nature of Dark Matter and the mechanisms after which cosmic rays are accelerated up to 10$^{20}$\,eV energies.

\begin{figure}
\centering
\includegraphics[trim=0.0cm 0.0cm 0.0cm 0.0cm, clip=true, width=0.49\textwidth]{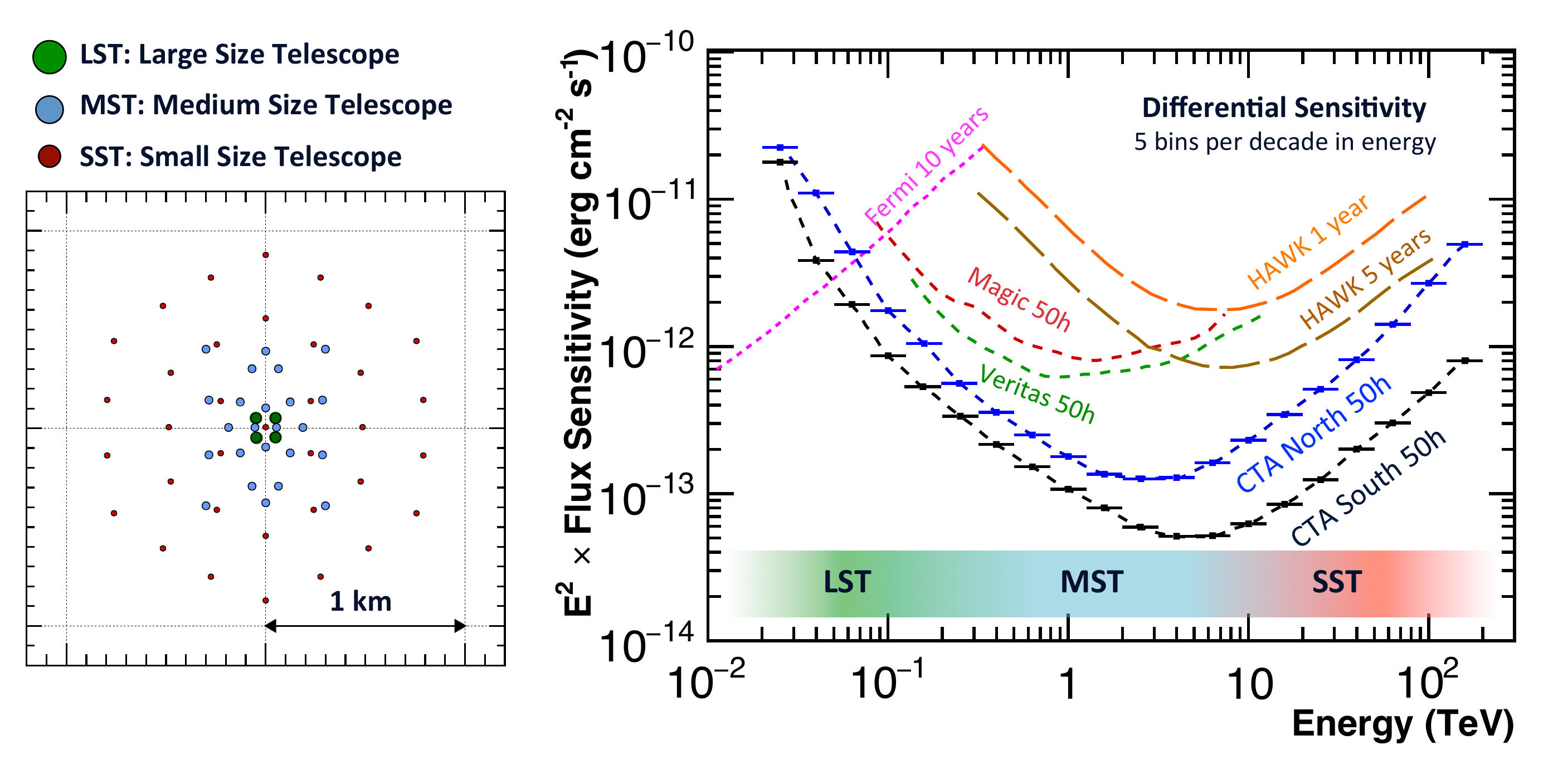}
\caption{Left: A possible configuration for the Southern hemisphere CTA array\,\cite{maier}. Right: Prediction for the sensitivity of CTA to a point-like gamma-ray source compared with the actual generation of ground telescopes and space detectors\,\cite{cta,hassan}. }
\label{Fig:cta}
\end{figure}

The  Cherenkov Telescope Array (CTA) Consortium is developing the new generation of  Imaging Atmospheric Cherenkov Telescopes (IACTs) detectors. IACTs detect the UV Cherenkov light produced in the atmospheric showers initiated by cosmic rays: the analysis of the Cherenkov light at ground allows to identify electromagnetic showers initiated by cosmic ray photons and $\mbox{e}^\pm$ with energies above 10\,GeV and to measure the energy and incoming direction of such cosmic rays.
The CTA collaboration aims to extend the flux sensitivity of the current IACT telescope generation by at least one order of magnitude in the core energy range and to extend the energy range beyond 100 TeV to explore the gamma-ray sky at the highest energy regime and below 100\,GeV to overlap with the current space-borne gamma-ray detectors.
To achieve this goal, the CTA collaboration will install telescopes with different acceptances and configurations in two arrays, one in southern and one in the northern hemisphere to maximize the coverage of the sky. (see Figure\,\ref{Fig:cta})  The CTA Observatory will be operated as a proposal-driven observatory, openly accessible to a large scientific community.

\section{FBK NUV-HD SIPMs}
Silicon PhotoMultipliers (SiPM) are arrays of microcells of Avalanche Photo Diodes (APD) connected in parallel and operated in Geiger mode. They are characterized by timing of the order of $\sim$\,ns, with recovery times comparable to those of conventional phototubes. Their gain factors of $\sim\,10^6$ make them proper detectors for the detection of single photons hitting their active surface. Their compactness and robustness allow to easily deploy SiPM sensors on wide area telescope focal planes. Their endurances to the night sky background light will allow to extend the telescope duty cycles acquiring quality data also during bright Moon nights. SiPM are therefore ideal sensors to equip the CTA telescope cameras.

\begin{figure}
\centering
\includegraphics[trim=0.0cm 0.0cm 0.0cm 0.0cm, clip=true, width=0.49\textwidth]{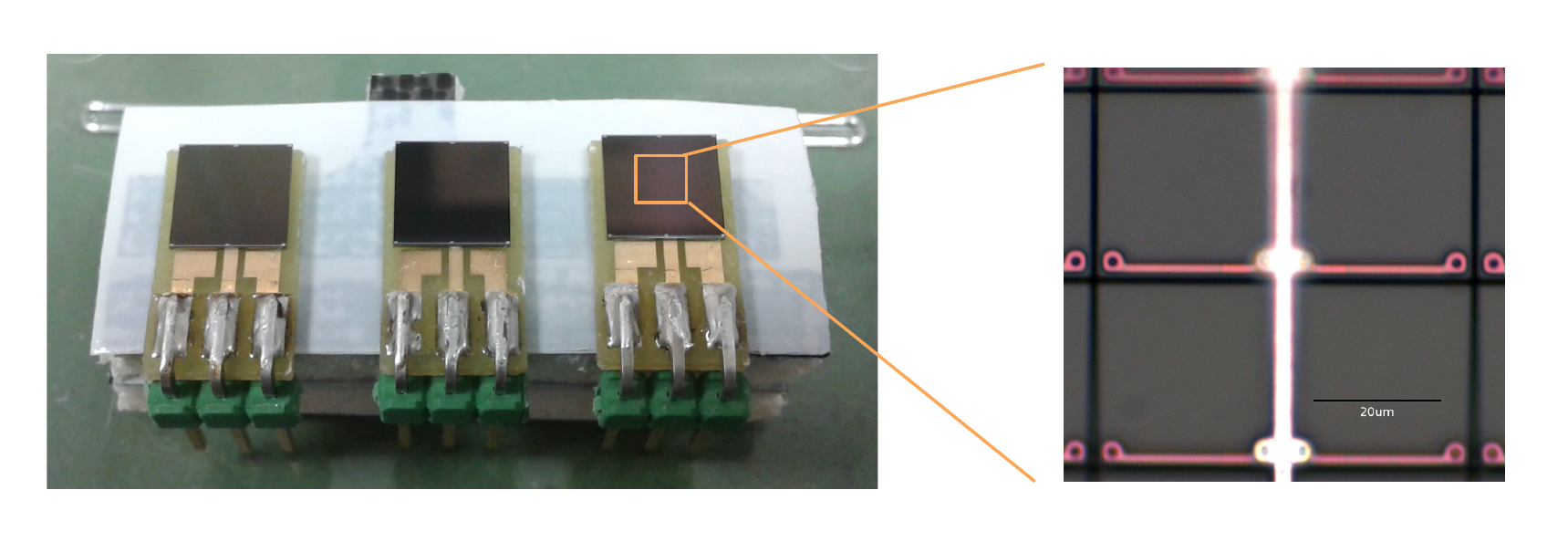}
\caption{Left: 6\,x\,6\,mm$^{2}$ NUV-HD SiPMs with microcell area of area 30\,x\,30\,$\mu\mbox{m}^2$ assembled for laboratory tests. Right: Zoom of 4 microcells. The metal connections and the quenching resistors are also visible.  The active area geometrical fill factor amounts to 76\%.}
\label{Fig:sipm}
\end{figure}

\begin{figure}
\centering
\hspace*{-0.50cm}
\includegraphics[trim=0.0cm 0.0cm 0.0cm 0.0cm, clip=true, width=0.56\textwidth]{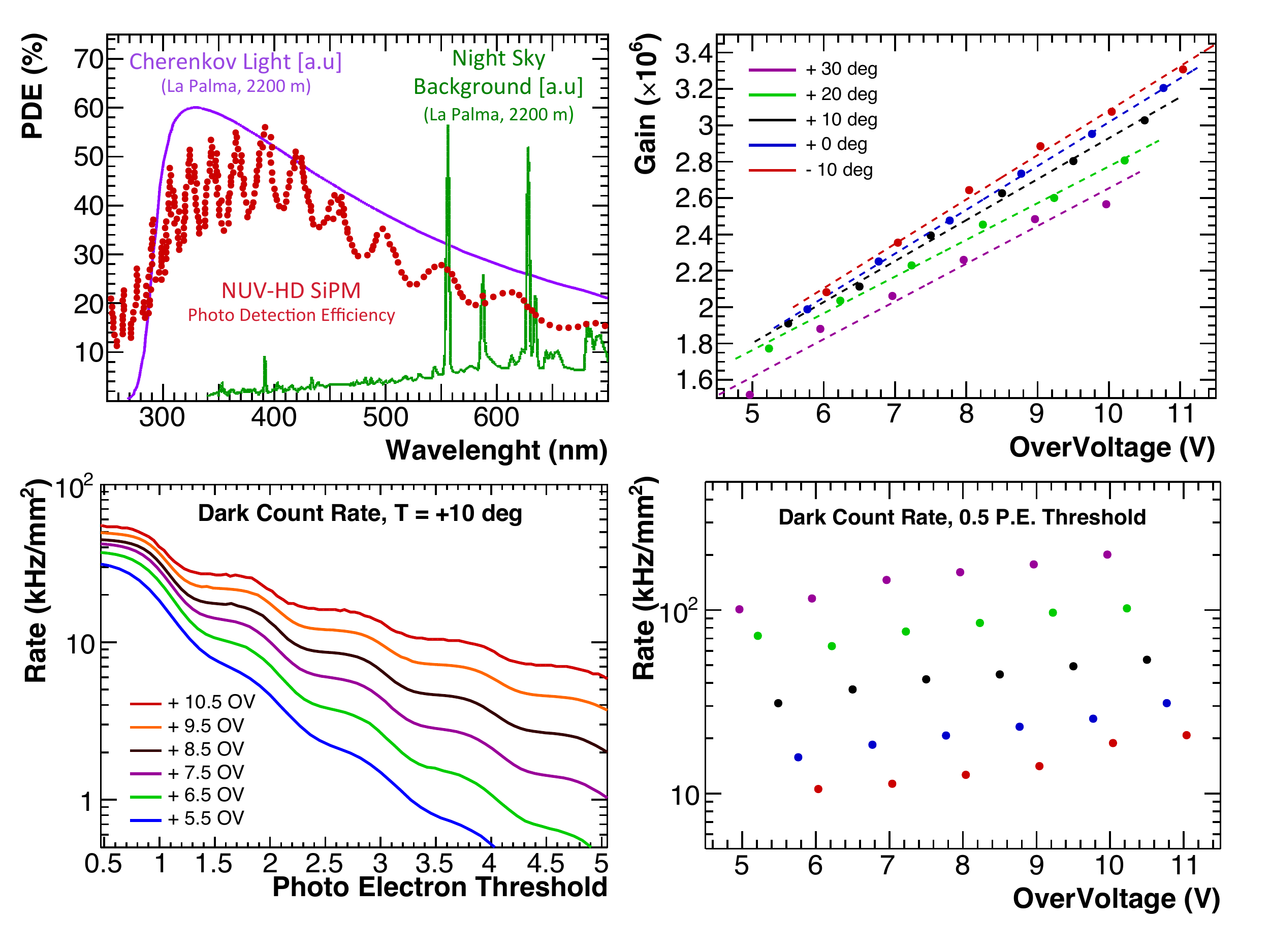}
\caption{Top left: Photo detection efficiency of the SiPM sensor\,\cite{nepomuk} superimposed with the expected spectra of the Cherenkov signal light and of the diffuse night sky background for one of the candidate CTA north sites. Top right: SiPM gain as function of operating voltage for several operating temperatures. Bottom left: Dark count rate measured for different operating voltages at temperature T=10$^{0}$C and as function of the p.e. threshold. Bottom right: Dark count rate for different operating voltages and temperatures with a threshold of 0.5 p.e.}
\label{Fig:perf}
\end{figure}

SiPM sensors have been developed by FBK\cite{fbk} in cooperation with INFN as a proposal to equip the focal plane of CTA telescopes.
The latest sensor technology is based on 6\,x\,6\,mm$^{2}$ p$^+$-n SiPMs with microcell area of 30\,x\,30\,$\mu\mbox{m}^2$, sensitive to UV light and with breakdown voltages of $\sim$\,28\,V, named High Density NUV (NUV-HD, Figure\,\ref{Fig:sipm}). The sensors have been characterized in the clean room and laboratory tests with pulsed UV sources and in dark conditions, using a transimpedence amplifier ASD-EP-EB-N developed by Advansid for the evaluation of the FBK sensor performances\,\cite{advansid}. The devices show competitive performances with respect to commercial devices, with a Photon Detection Efficiency (PDE) better than 40\% in the expected UV signal range, single photoelectron noise rate below 100\,kHz/mm$^2$ at operating temperature and an optimal signal/noise for single photoelectron detection (Figure\,\ref{Fig:perf}).

\section{Development and assembly of a possible solution for the upgrade of the pSCT telescope camera}
\begin{figure}
\centering
\includegraphics[trim=0.0cm 0.0cm 0.0cm 0.0cm, clip=true, width=0.47\textwidth]{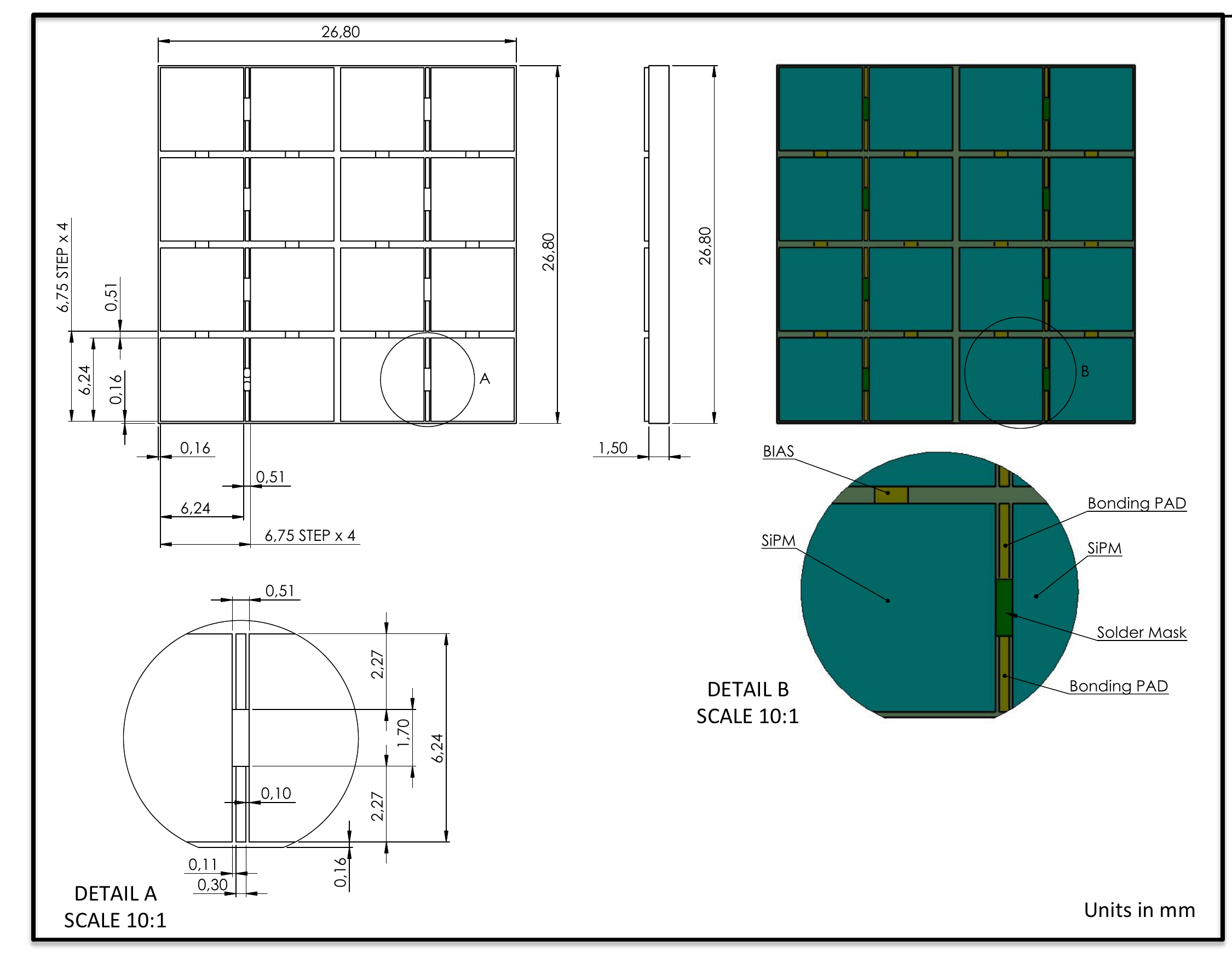}
\caption{Design of the PCB corresponding to 1 unit hosting 16 SiPMs. The sensors are arranged to achieve a uniform coverage of the camera area. 16 bonding pads are placed along the two corridors between two lines of sensors to couple the SiPM anodes to the back of the PCB.}
\label{Fig:pcb}
\end{figure}

The NUV-HD SiPM sensors described in the previous section have been selected to assemble a possible upgrade of the focal plane camera of the Prototype Schwarzschild-Couder Telescope (pSCT)\,\cite{psct}, a CTA Medium Size Telescope (MST) prototype that will be operated at the Fred Lawrence Whipple Laboratory in Arizona (US) in 2017. The current camera design is based on Hamamatsu S12642-0404PA-50(X) MPPC modules. The improved sensitivity measured for the NUV-HD SiPM sensors has driven the efforts to integrate this technology to the pSCT camera. INFN together with Italian universities has designed and developed multi-SiPM modules compatible with the already existing pSCT camera mechanics and readout. Each module is made of four independent units (or quadrants), each equipped with 16 NUV-HD SiPMs. The units have been designed to cover a 26.9\,x\,26.9\,mm$^2$ area, corresponding to 53.8\,x\,53.8\,mm$^2$ area of each module. Each 64 SiPM module will be arranged in a chessboard geometry with a module-module distance of 200\,$\mu$m. The mechanical layout of the 16 SiPM PCB is shown in Figure\,\ref{Fig:pcb}. With this configuration, the module pitch is 54\,mm, while the sensor pitch is 6.75\,mm, equally distributed all over the camera area. The nominal distance between sensor edges amounts to 510\,$\mu$m. The bias voltage is applied at the back common cathode of all the 16 sensors, while the signal of each sensor is individually readout through a wire-bond connection between a conductive pad placed on the sensor edges and a bonding pad placed on the PCB and in conductive connection with the back of the PCB.  

\begin{figure}
\centering
\includegraphics[trim=0.0cm 0.0cm 0.0cm 0.0cm, clip=true, width=0.49\textwidth]{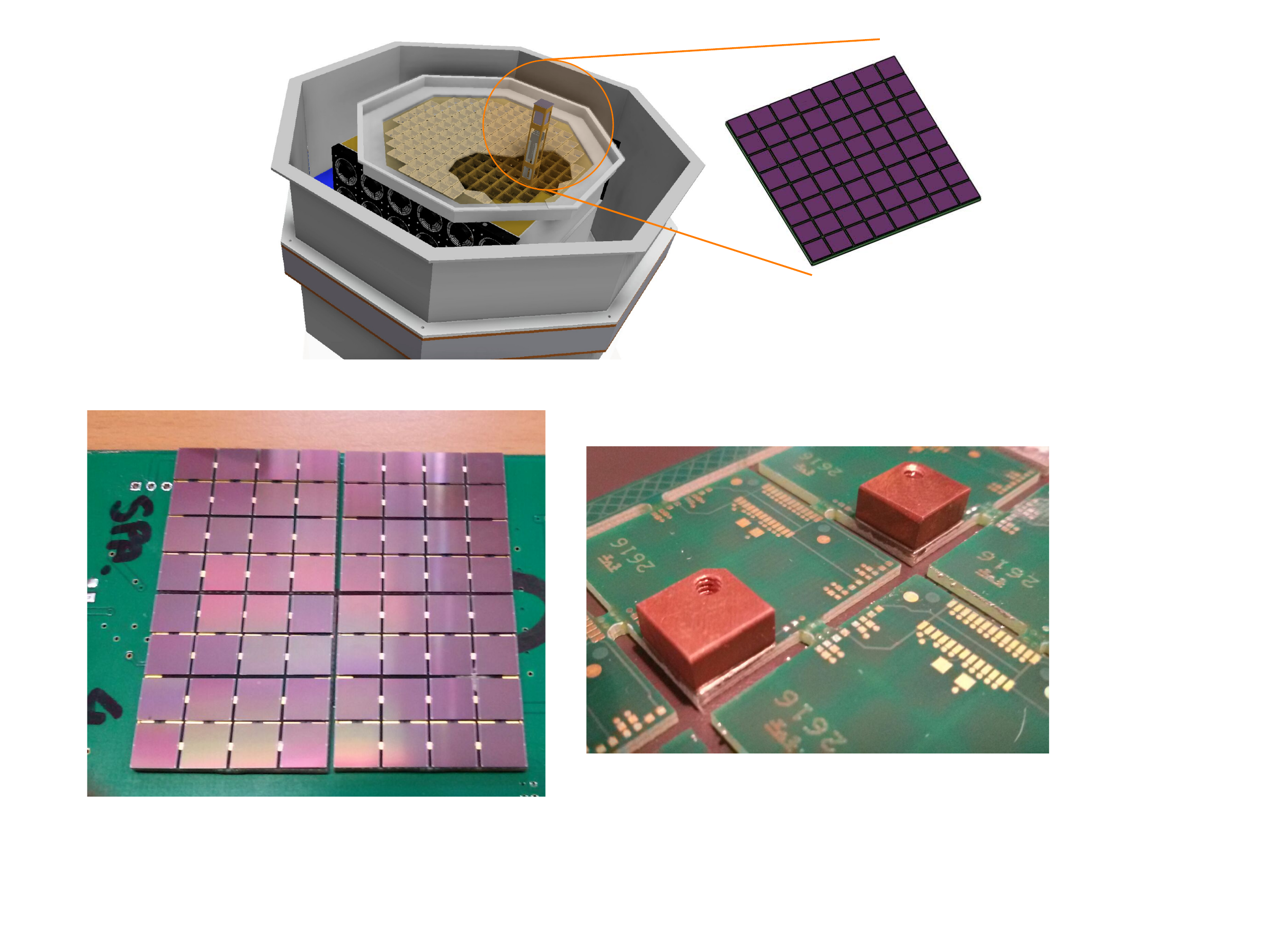}
\caption{Top: Conceptual scheme of the pSCT telescope focal plane camera with one module equipped with 64 NUV-HD SIPMs. Bottom left: 4 units assembled with 16 SiPMs. Bottom right: back of the PCBs assembled with the copper block.}
\label{Fig:matrix}
\end{figure}

\begin{figure}
\centering
\includegraphics[trim=0.0cm 0.0cm 0.0cm 0.0cm, clip=true, width=0.49\textwidth]{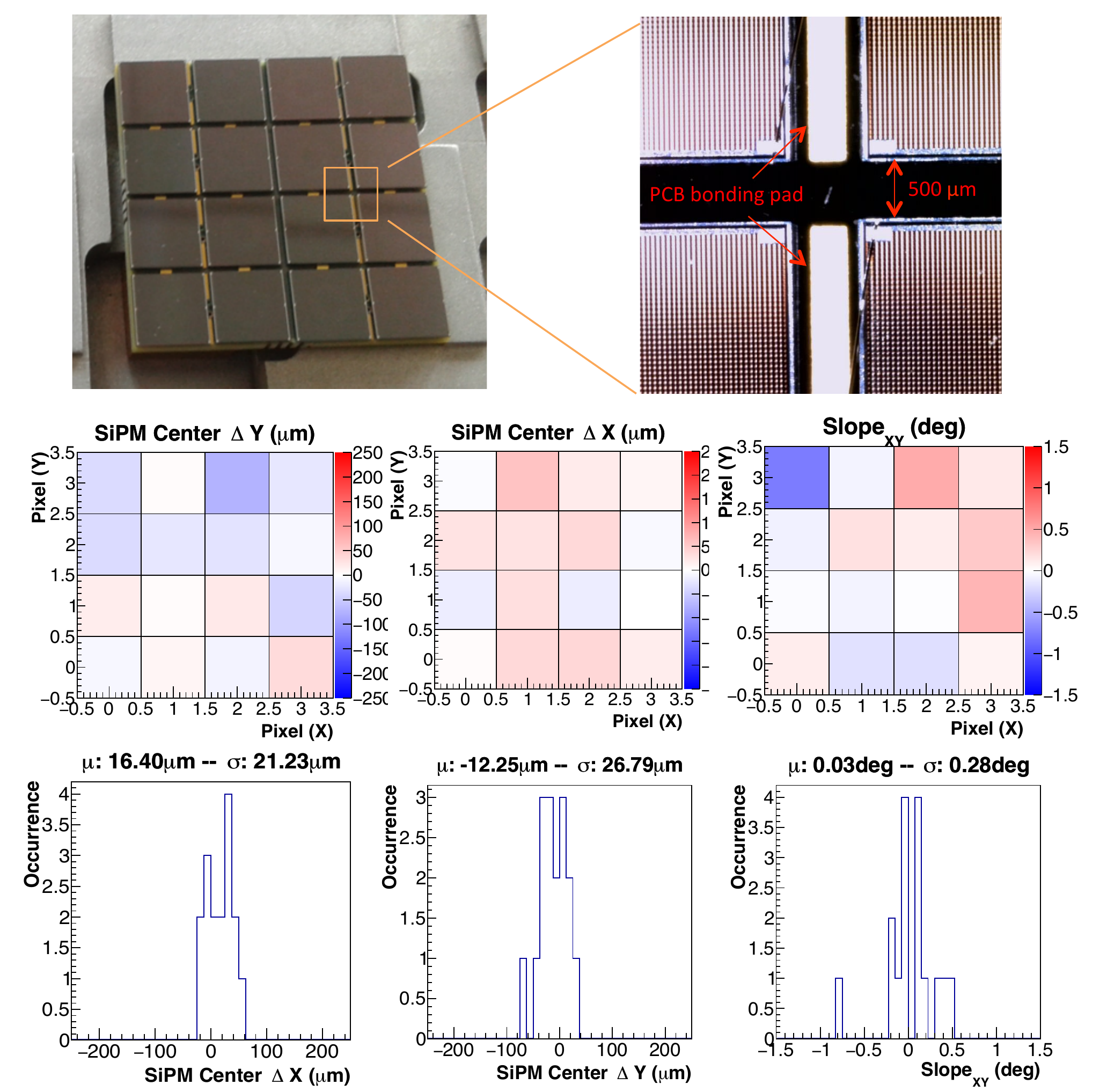}
\caption{Top: 1 unit with 16 NUV-HD SiPMs assembled with a pick \& place machine. The zoom on the right highlights the sensor corners and the PCB binding pads. Bottom: Sensor misalignment from nominal position for one unit.}
\label{Fig:alignment}
\end{figure}

Before the placement of SiPMs, the PCB are thermally coupled to the copper blocks used as mechanical structure of the PCB and to radiate excess heat produced by the electronics. This step is crucial to achieve the best possible alignment of the modules when assembled in the camera mechanics. Custom mechanical holders with holes and position pins allow to achieve a 50\,$\mu$m accuracy for the alignment in the xy plane. 
A ``pick \& place" machine is used to dispense the conductive glue to the PCB top layer metal pads and place the SiPM sensors on the PCB. Some examples of the products of assembly tests are shown in Figure\,\ref{Fig:matrix}.
Preliminary tests to validate the accuracy of the pick \& place machine have shown a sensor alignment better than $40\,\mu\mbox{m}$ and a board flatness better than $80\,\mu\mbox{m}$, well within the requirements (Figure\,\ref{Fig:alignment}).

\begin{figure}
\centering
\includegraphics[trim=0.0cm 0.0cm 0.0cm 0.0cm, clip=true, width=0.47\textwidth]{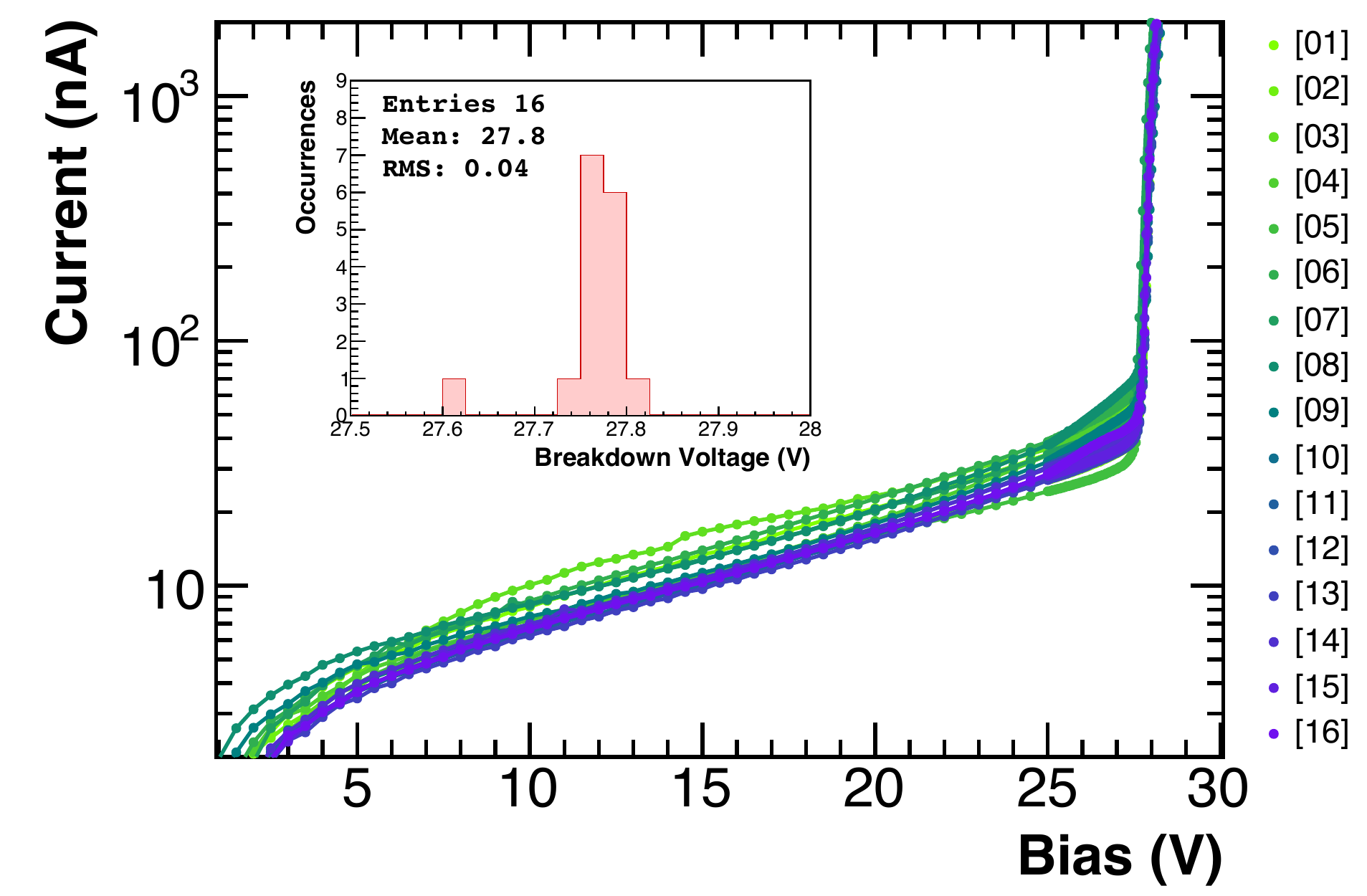}
\caption{Characteristics current-voltage (IV) curves for 16 SiPMs on the same board measured at room temperature. Insert: distribution of breakdown voltages for the same 16 SiPMs. The sensors show a uniformity better than 0.5\% in the breakdown conditions.}
\label{fig:iv}
\end{figure}

After the SiPM placement, the 16 sensor bonding pads are wire bonded to the PCB bonding pads and then tested to check the quality of the conductive connections. The characteristic current-voltage (IV) curves of each sensor of the unit are measured in series using a custom switching matrix with digital control of the multiplexer. An example of IV curves for 16 sensors of one unit is shown in Figure\,\ref{fig:iv}. \\
A layer of UV transparent epoxy is finally dispensed over the SiPM to protect the wire bonds and the sensor surface. Since the UV light absorption probability increases with depth, the epoxy layer has to be dispensed with the highest possible level of uniformity over all the module, avoiding border effects. Preliminary tests of epoxy deposition and transmittance are currently ongoing, with encouraging prospectives.


\section{Test and validation campaign}
To validate the camera concept, 9 quadrants of the pSCT camera will be initially equipped with the camera solution described in the previous section. A total of 100 modules, corresponding to 1600 SiPM sensors, will be tested in the INFN laboratories and a selection of 36 modules will be mounted on the pSCT camera. A DAQ system based on a 16-channel charge amplifier has been developed for this purpose. A Pole-Zero cancellation circuit, optimized to achieve a 20 ns pulse with an amplitude of 4\,mV/p.e. and to minimize the electronics signal offset, is applied to reshape the signal\,\cite{cbonavolonta}. The output signal of the preamplifier board is subsequently readout with a V792 QDC module.

Preliminary tests of the preamplifier performances have been performed on a 16-channel prototype unit using a pulsed 380\,nm light\,\cite{dsimone}. The response of each single sensor has been studied by covering all other channel with a custom mask. The results show an evident quantization of the signal down to the single photoelectron, with a signal/noise ratio of $\sim$\,5.

\begin{figure}
\centering
\includegraphics[trim=0.0cm 0.0cm 0.0cm 0.0cm, clip=true, width=0.49\textwidth]{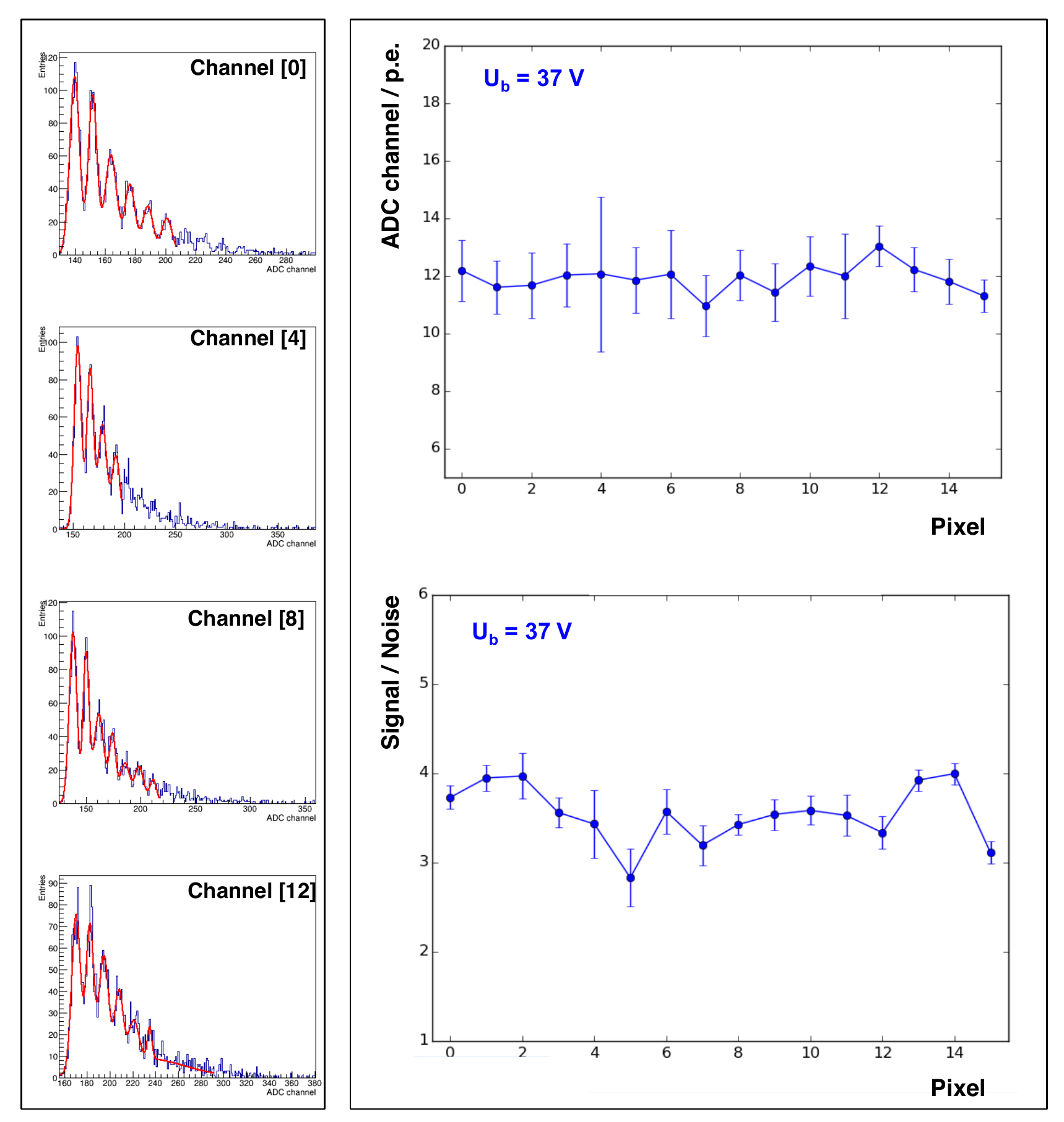}
\caption{Left: ADC spectra for 4 sensors of the 16-channel unit measured with the DAQ system based on a QDC charge integrator matrix with a 100\,ns integration time window. Right: Gain and signal/noise ratio for the 16 channels of one unit measured with the same setup.}
\label{fig:16chan}
\end{figure}

Subsequently, the whole DAQ system has been tested illuminating the whole unit and acquiring the 16 signals with the QDC module. The signal is integrated over a time window of 100\,ns. Each ADC channel corresponds to a charge of $\sim$\,100\,pC. The results show a good uniformity in the response of the 16-channel unit, with a slightly lower gain and signal/noise ratio probably introduced by the intrinsic impedences of the setup (see Figure\,\ref{fig:16chan}). Further efforts are ongoing to optimize and calibrate the DAQ setup towards the massive 1600 sensor tests.

\section{Outlook}
SiPM sensors present suitable properties to equip the focal planes of Imaging Air Cherenkov Telescope cameras, allowing to improve the camera performance with respect to the use of standard photomultipliers. The latest technology achieved in the SiPM sensors produced by FBK has driven the design of a possible upgrade of the focal plane camera of the pSCT CTA telescope, whose current design is using an older generation of Hamamatsu MPPC. Modules made of 16 SiPM sensors have been designed to be compatible with the current camera mechanics and readout, and are currently being assembled. 
Preliminary tests and verifications have confirmed the quality of the assembly and of the performance of the modules. In 2017 at least 36 units of 16 sensors each will be ready for integration with the pSCT telescope camera and tested on site.

%




\bigskip 
\begin{acknowledgments}
The authors gratefully acknowledge support from agencies and organizations under Funding Agencies at www.cta-observatory.org.
\end{acknowledgments}

\bigskip 

\end{document}